\documentclass[10pt,final,doublecolumn]{IEEEtran}
\IEEEoverridecommandlockouts
\usepackage{multirow}
\usepackage{amsmath}
\usepackage{amssymb}
\usepackage{amsthm}
\usepackage{mathrsfs}
\usepackage{latexsym}
\usepackage{graphicx}
\usepackage{bbding}
\usepackage{indentfirst}
\usepackage{cases}
\usepackage{supertabular}
\usepackage{algorithm,algorithmic}
\usepackage{subeqnarray}
\usepackage{color}
\usepackage{bm}
\usepackage{stfloats}
\usepackage{subfigure}
\usepackage{array}
\usepackage{algorithm,float}

\IEEEoverridecommandlockouts
\allowdisplaybreaks[4]



\begin{document}
\title{LEO Satellite Internet of Things: Architecture, Technology, and On-Orbit Verification}
\author{\IEEEauthorblockN{Ming Ying, Xiaoming Chen, Qiao Qi, Yichao Xu, and Jiajun Pan}
\thanks{Ming Ying, Xiaoming Chen, Yichao Xu and Jiajun Pan are with the College of Information Science and Electronic Engineering, Zhejiang University, Hangzhou 310027, China (e-mail:\{ming\_ying, chen\_xiaoming, yichao\_xu, jiajun\_pan\}@zju.edu.cn). Qiao Qi is with the School of Information Science and Technology, Hangzhou Normal University, Hangzhou 311121, China (email: qiqiao@hznu.edu.cn). 
}}\maketitle

\begin{abstract}
   Low Earth orbit (LEO) satellite constellations are poised to become a cornerstone of the sixth-generation (6G) Internet of Things (IoT), providing truly global coverage and ubiquitous connectivity. This article presents a holistic two-dimensional system architecture for 6G LEO satellite IoT that incorporates composition and functional perspectives to facilitate the seamless integration of LEO satellites and terrestrial networks. Building upon this architecture, we evaluate three pivotal enabling technologies targeting the uplink, downlink, and inter-satellite links (ISLs). Specifically, we analyze massive grant-free random access for efficient uplink connectivity, investigate deep learning-based multibeam precoding for robust downlink transmission, and examine distributed cooperative routing for resilient ISL data delivery. Furthermore, we present an on-orbit verification platform that validates the real-world feasibility and performance of the proposed solutions. Finally, we outline key open challenges and future research directions to guide the realization of future LEO satellite IoT. 
\end{abstract}

\begin{IEEEkeywords}
6G, low Earth orbit satellite, Internet of Things, on-orbit verification
\end{IEEEkeywords}

\section{Introduction}
	The evolution toward sixth-generation (6G) wireless networks is driven by the vision of seamless, intelligent, and global connectivity \cite{r1}, with massive Internet of Things (IoT) support serving as a key requirement for applications such as environmental sensing, precision agriculture, and maritime logistics. Because these services are often deployed in remote areas beyond terrestrial infrastructure, universal coverage requires a unified architecture that integrates ground networks with space-based systems \cite{r2, r3}. In this context, low Earth orbit (LEO) satellite IoT systems have emerged as a critical enabler of future 6G networks by providing ubiquitous, infrastructure-independent, and resilient connectivity, particularly when terrestrial networks are disrupted by natural disasters \cite{r4}-\cite{r6}. Moreover, with high-capacity transmission and advanced beamforming techniques, LEO constellations can support large-scale concurrent connections for both stationary and highly mobile platforms. Recognizing these advantages, the 3rd Generation Partnership Project (3GPP) has progressively incorporated non-terrestrial networks (NTN) into its standardization roadmap, from exploratory studies in Release 15 to normative 5G NTN specifications in Release 17 \cite{r7,r8}, while ongoing work in Releases 18 and 19 further targets massive IoT optimization, including enhanced uplink capacity, regenerative payloads, and reduced-capability device connectivity \cite{r9}. These standardization efforts lay an essential foundation for global interoperability and the seamless integration of LEO satellites into 6G networks.
	
	While these advancements are promising, the large-scale deployment of high-performance LEO satellite IoT faces a series of complex, interconnected technical obstacles which existing specifications cannot be addressed \cite{r10}-\cite{r12}. Firstly, LEO satellite IoT must accommodate massive, sporadic uplink access from low-power IoT devices, a task complicated by high Doppler shifts and stringent energy constraints. Secondly, delivering robust and spectrally efficient downlink transmission via multibeam antennas remains difficult due to rapidly changing channels and imperfect channel state information (CSI). Finally, the highly dynamic topology of emerging mega LEO satellite constellations requires innovative routing protocols that ensure low-latency data delivery while minimizing signaling overhead. Addressing these interconnected challenges is critical to achieving the high levels of reliability and scalability required for the 6G era.
	
	Motivated by these challenges, this article provides a comprehensive investigation into the system architecture and enabling technologies of LEO satellite IoT within the 6G framework. We begin by proposing a holistic two-dimensional system architecture, encompassing composition and functional perspectives. Building on this foundation, we offer an in-depth evaluation of fundamental enabling technologies, including grant-free random access (GF-RA) schemes, deep learning (DL)-based multibeam precoding techniques, and distributed cooperative routing strategies. Furthermore, to bridge the gap between theoretical models and practical implementation, we present an on-orbit verification platform that validates the real-world feasibility and performance of the proposed solutions. Finally, we present several unresolved research challenges and future directions for 6G LEO satellite IoT. {The comparison of the proposed architecture with existing LEO satellite IoT architectures are summarized in Table I.}
	
	\begin{table*}[!t]
		\small
		\centering
		\renewcommand{\arraystretch}{1.2}
		\caption{Comparison of the Proposed architecture with Existing LEO Satellite IoT Architectures}\label{tab:Novelty_Comparison}
		\begin{tabular}{|>{\centering\arraybackslash}p{4cm}|>{\centering\arraybackslash}p{4.2cm}|>{\centering\arraybackslash}p{5cm}|}
			\hline
			\textbf{Features} & \textbf{Existing Architectures} & \textbf{Proposed Architectures} \\
			\hline\hline
		\textbf{System Architecture} & Fragmented design and ``bent-pipe'' relays & \textbf{Holistic two-dimensional (2D)} architecture \\
			\hline
			\textbf{Uplink Access Scheme} & GB-RA or OMA & \textbf{Tensor-based} GF-RA \\
			\hline
			\textbf{Downlink Precoding} & Statistic models within perfect CSI & \textbf{DL-based} precoding within imperfect CSI \\
			\hline
			\textbf{ISL Routing Protocol} & Centralized routing & \textbf{Distributed cooperative} routing  \\
			\hline
			\textbf{Validation Method} & Purely software simulations & \textbf{Real-world on-orbit} verification \\
			\hline
		\end{tabular}
	\end{table*}
	
	The remainder of this article is organized as follows. Section II details the proposed two-dimensional system architecture. Section III analyzes key enabling technologies, namely massive grant-free random access, DL-based multibeam precoding, and distributed cooperative routing. Section IV validates these solutions via an on-orbit verification platform, while Section V outlines open challenges and future research directions. Finally, Section VI concludes the article.  

    \section{System Architecture for LEO satellite IoT}
   	In this section, we propose a holistic, two-dimensional system architecture for LEO satellite IoT, encompassing both composition and functional architectures. By clearly delineating the physical infrastructure from its core operational functions, this approach fundamentally decouples hardware deployment from intelligent service orchestration. Ultimately, this architecture design establishes a resilient, scalable, and highly efficient technical reference for realizing the next-generation LEO satellite IoT.
    \begin{figure*}[h]
    	\centering
   		\includegraphics [width=0.9\textwidth] {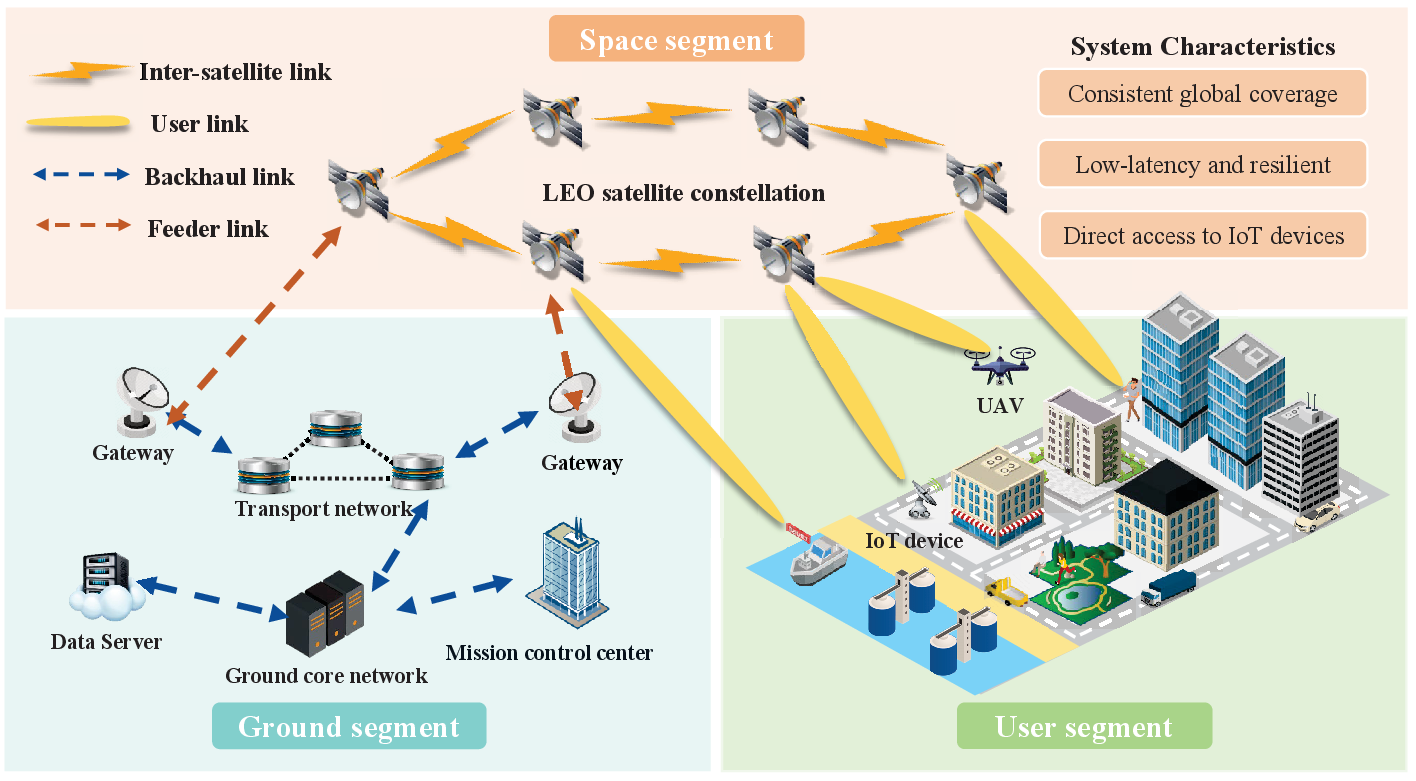}
    	\caption {Composition architecture of LEO satellite IoT. {The user link denotes the direct access link between IoT devices and LEO satellites, the inter-satellite link denotes the link between neighboring satellites for intra-constellation forwarding, the feeder link denotes the satellite-to-gateway link, and the backhaul link denotes the terrestrial connection from gateways to the core network and data servers.}} \label{Fig2-1}
    \end{figure*}
    
    \subsection{Composition Architecture}
    Building upon the two-dimensional framework introduced above, we first detail the composition architecture for LEO satellite IoT. Specifically, composition architecture partitions the whole system into three core segments: the space, user, and ground segments, as depicted in Fig. \ref{Fig2-1}.
    
    {For the space segment, it consists of LEO satellites and inter-satellite links (ISLs), thereby forming the backbone transmission network in space\footnote{The considered LEO satellite IoT adopts an asymmetric frequency division duplex (FDD) architecture, with L-band for uplink and Ku-band for downlink. This is because L-band offers lower path loss and better robustness for low-power uplinks, while Ku-band provides wide bandwidth for high-capacity downlinks.}. Organized symmetrically to ensure continuous global coverage, these satellites transcend traditional "bent-pipe" relays to operate as intelligent orbital base stations providing direct IoT access. To satisfy the massive 6G capacity demands, modern constellations frequently adopt multi-layer architectures, deploying LEO satellites across varying altitudes and orbital inclinations. By coupling with optical ISLs, the space segment enables autonomous, intra-constellation data routing, significantly reducing transmission latency and minimizing reliance on ground infrastructure to ensure resilient global service.
       
   	The user segment comprises massive, heterogeneous IoT devices serving as the ultimate endpoints for data generation and service execution. A defining hallmark of this 6G-oriented architecture is its fundamental shift toward direct-to-satellite (D2S) connectivity. By bypassing the physical limitations of traditional terrestrial cellular infrastructure, this paradigm enables IoT devices to establish ubiquitous, direct links with LEO satellite constellations. Consequently, diverse IoT devices ranging from low-power environmental sensors to autonomous industrial equipment can remain seamlessly integrated into the global network, regardless of their geographic isolation. By tightly coupling these nodes to the spatial backbone, the user segment efficiently translates raw space-ground connectivity into pervasive, real-time IoT applications.
    
    The ground segment provides the essential terrestrial foundation for LEO satellite IoT, integrating gateway stations, the transport network, the ground core network, and the mission control center. Within this architecture, the gateway stations serve as the primary interface, facilitating the exchange of data between the LEO satellite constellations and ground-based systems. Simultaneously, the mission control center provides the necessary oversight to manage the operational health and orbital positioning of the satellites. By combining these hardware and software components, the ground segment provides the essential control and connectivity required to maintain a stable and responsive global communication architecture.
    
    Moreover, the end-to-end data flow includes uplink and downlink procedures. In the uplink, IoT devices transmit data to visible LEO satellites through user links, and the data is then forwarded through ISLs or delivered to gateway stations via feeder links. In the downlink direction, control messages, service data, or task instructions are generated by the ground network or application servers, delivered to the gateway station, forwarded to the satellite constellation, and finally transmitted to IoT devices.
    
    Finally, the proposed architecture assumes D2S IoT access, ISL-enabled data forwarding, and time-varying network topology caused by satellite mobility. These assumptions define the scope of the proposed architecture and provide the basis for the enabling technologies discussed in Section III.
    }

    \subsection{Functional Architecture}

	\begin{figure*}[h]
		\centering
		\includegraphics [width=0.9\textwidth] {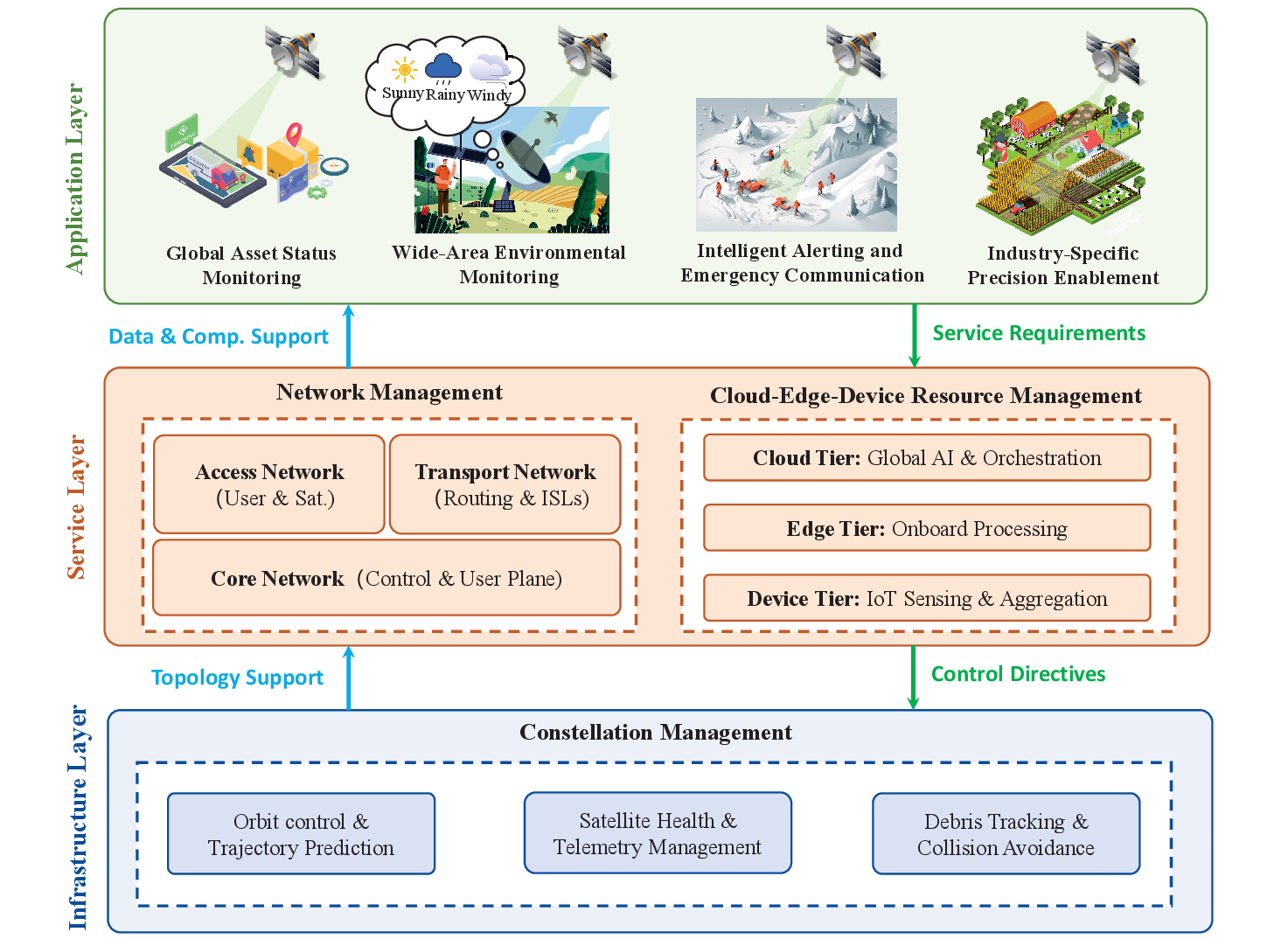}
		\caption {Functional architecture of LEO satellite IoT.} \label{Fig2-3}
	\end{figure*}
	
Building upon the physical segments in the composition architecture, the functional architecture is organized into a hierarchical three-tier framework, i.e., the infrastructure layer, the service layer, and the application layer, as illustrated in Fig. 2. The subsequent discussion details the specific roles and mechanisms defining each layer.

\textbf{\textit{1. Infrastructure Layer:}} Establishing the fundamental physical and spatial basis of the LEO satellite IoT system, this layer is dedicated exclusively to the operational integrity of space assets via \textbf{Constellation Management}. Its primary mandate is the precise regulation of orbital paths and position-keeping to maintain the specific geometric topology required for seamless global coverage. To achieve this, the system continuously monitors the telemetry, health, and trajectory of each satellite, facilitating accurate orbital predictions essential for network-wide coordination. 

\textbf{\textit{2. Service Layer:}} Operating as the intelligent middleware of the architecture, the service layer bridges the physical infrastructure with high-level user applications. It provides the ubiquitous connectivity and distributed computational power necessary to process massive IoT data volumes, comprising two core pillars in the following:
\begin{itemize}
	\item \textbf{Network Management:} This module orchestrates end-to-end data connectivity and routing throughout the system. First, the access network synchronizes user-satellite links, managing massive IoT device registrations while ensuring seamless handovers across rapidly moving footprints. Second, the transport network executes control-plane management to dynamically optimize ISLs and establish resilient space-ground backhaul routing. Finally, the core network oversees user-plane operations, processing actual data payloads, coordinating sessions, and handling service accounting to maintain a secure, continuous communication loop.
	\item \textbf{Cloud-Edge-Device Resource Management:} This function coordinates distributed computing and storage resources across the cloud, edge, and device tiers to meet the processing demands of 6G IoT. Firstly, the \textbf{cloud tier} provides centralized artificial intelligence (AI) processing, global orchestration, and large-scale data analytics through ground-based data centers. In the middle, the \textbf{edge tier} integrates mobile edge computing (MEC) capabilities into LEO satellites and terrestrial gateways to enable real-time processing, reduce latency, and alleviate space-ground backhaul congestion. Finally, the \textbf{device tier} encompasses diverse IoT devices responsible for ubiquitous sensing and localized data aggregation. Together, this cloud-edge-device continuum enables efficient resource utilization and distributed intelligence across LEO satellite IoT systems.

\end{itemize}

\textbf{\textit{3. Application Layer:}} By leveraging the connectivity and computational resources of the service layer, the application layer translates raw data into actionable insights through specialized services tailored to diverse requirements. This layer encompasses four primary functional domains as follows
\begin{itemize}
	\item \textbf{Global Asset Status Monitoring:} This service facilitates the continuous tracking and status evaluation of assets worldwide. Key scenarios include the real-time tracking of ocean freight containers to optimize supply chains, and global flight and vessel surveillance to enhance maritime and aviation safety. Additionally, it supports scientific efforts such as wildlife migration research and provides remote monitoring for critical industrial equipment in isolated regions.
	\item \textbf{Wide-Area Environmental and Infrastructure Monitoring:} This application enables persistent data collection across vast, often inaccessible geographic areas. It facilitates automated weather station data collection and geological hazard monitoring to detect risks like landslides. Furthermore, it supports industrial safety through pipeline inspections and contributes to climate science by enabling polar and glacier research in extreme environments.
	\item \textbf{Intelligent Alerting and Emergency Communication:} This functionality establishes a resilient communication lifeline during crises when traditional terrestrial networks fail. It is designed to process distress beacon signals for search and rescue operations, and to broadcast earthquake and tsunami early warnings to vulnerable populations. It also enables automated forest fire alerts and ensures emergency command connectivity for first responders in outage areas.
	\item \textbf{Industry-Specific Precision Empowerment:} This service provides targeted data insights to optimize the efficiency and safety of specialized sectors. It supports precision agriculture by monitoring soil and crop conditions, and enhances smart forestry through automated management services. Additionally, it improves industrial safety via mine stability monitoring, delivering the high-resolution data necessary to protect personnel in hazardous environments.
\end{itemize}

By decoupling physical infrastructure, data services, and application delivery, this hierarchical three-tier architecture establishes a resilient foundation for dynamic industry applications. However, translating this theoretical vision into a high-performance reality within a harsh orbital environment requires the advanced technological breakthroughs, which are examined in the subsequent section.

	\section{Key Technologies of LEO satellite IoT}
	In this section, we discuss the key enabling technologies required to translate the proposed architecture into a high-performance reality. Specifically, we examine fundamental innovations across the space-ground uplink, downlink, and intra-constellation ISLs to overcome the inherent challenges of massive IoT integration and dynamic orbital topologies.
	
	\subsection{Massive Grant-Free Random Access}
	Efficient uplink access is paramount for massive connectivity in LEO satellite IoT. Traditional orthogonal multiple access (OMA) technologies, such as frequency division multiple access (FDMA), suffer from rigid resource allocation and poor scalability, making them ill-suited for bursty IoT traffic. Meanwhile, grant-based random access (GB-RA) incurs severe signaling overhead and latency due to request-grant handshakes, which is a critical flaw given the brief link availability of fast-moving satellites. To overcome these constraints, GF-RA provides a pivotal solution by enabling immediate data transmission without scheduling grants, thereby eliminating coordination delays. Furthermore, advanced GF-RA schemes leverage sophisticated multi-user detection to resolve uncoordinated packet collisions, drastically improving spectral efficiency to reliably accommodate massive scalability demands.
	
\begin{figure}[h]
		\centering
		\includegraphics [width=0.5\textwidth] {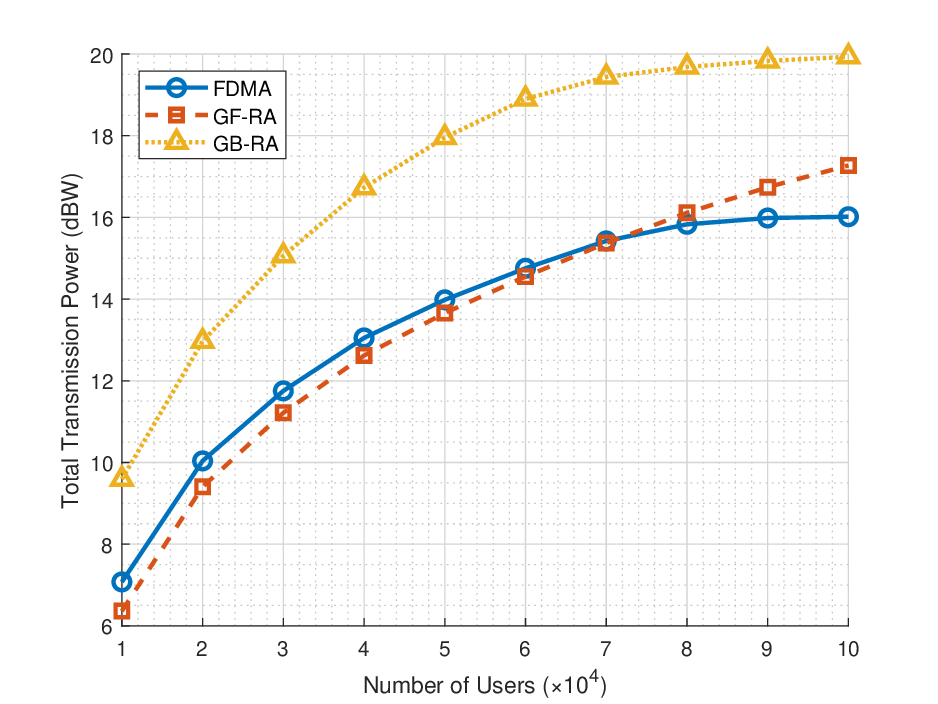}
		\caption {The total transmission power performance under different massive access technologies.} \label{Fig3-1}
	\end{figure}

	Building on the theoretical benefits of GF-RA, we previously introduced a specialized scheme in \cite{ma1} that represents received signals using a multi-dimensional tensor decomposition. This framework utilizes an efficient Bayesian learning algorithm to simultaneously perform device detection and channel estimation with low computational complexity. In this article, we evaluate this design in a high-quantity environment featuring up to 100,000 devices with the activity probability 0.1\%. {For uplink massive access, total transmission power is used to evaluate the energy efficiency of different access schemes under massive IoT connectivity and we perform 1000 trials to avoid uncertainties.} As illustrated in Fig. \ref{Fig3-1}, the GF-RA scheme matches the low power consumption of GB-RA while providing significantly higher data capacity. Furthermore, compared to FDMA, it facilitates superior frequency reuse by eliminating the need for dedicated channel assignments. The results demonstrate that the design effectively balances energy conservation with high transmission efficiency, making it a pivotal solution for low-power 6G IoT services.

	\subsection{Deep Learning-based Multibeam Precoding}
	Multibeam precoding plays a pivotal role in LEO satellite IoT by enabling directed, high-gain transmission toward massive IoT devices while suppressing inter-beam interference. However, traditional precoding schemes rely on explicit mathematical models that demand highly accurate CSI. In the highly dynamic LEO channel environment, rapid satellite mobility inevitably leads to delayed or imperfect CSI, causing severe performance degradation in conventional methods. To overcome this, DL provides a robust data-driven alternative. By leveraging historical CSI to implicitly capture complex channel dynamics, DL approaches dynamically optimize beamforming weights and compensate for estimation inaccuracies. Ultimately, shifting to these adaptive designs ensures resilient, high-capacity downlink connectivity even without instantaneous CSI.

	 \begin{figure}[h]
		\centering
		\includegraphics [width=0.5\textwidth] {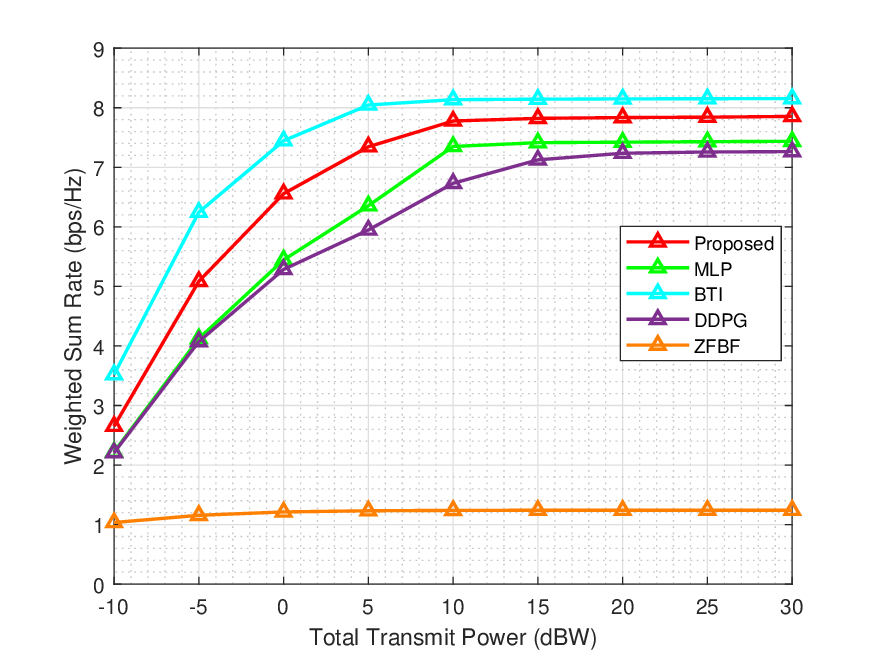}
		\caption {The weighted sum rate performance comparisons of different multibeam precoding schemes under various total transmit power.} \label{Fig3-2}
	\end{figure}
	
	Building on this, the authors in \cite{dl1} developed a DL-based framework to manage channel prediction and multibeam precoding in LEO satellite IoT. This architecture employs a CNN-LSTM framework to forecast CSI sequences, effectively neutralizing the signal fluctuations caused by rapid satellite movement. Furthermore, a VAE-based module is used to enhance the system's resistance to data inaccuracies. {For downlink multibeam transmission, weighted sum rate is used to evaluate spectral efficiency and throughput performance under different transmit power levels. The performance is obtained by averaging over an independent test set consisting of 1,000 channel realizations, thereby reducing the impact of random fluctuations.} As illustrated in Fig. \ref{Fig3-2}, the proposed scheme was evaluated against several benchmarks, including Bernstein-type inequality (BTI), zero-forcing beamforming (ZFBF), multi-layer perceptron (MLP), and deep deterministic policy gradient (DDPG). The results indicate that the BTI algorithm provides a near-optimal theoretical limit. However, its excessive computational demands resulted by iterative algorithms make it unsuitable for hardware deployment in space. In contrast, the DL-based scheme achieves comparable performance with significantly lower processing overhead, establishing it as a practical and efficient solution for the dynamic LEO satellite IoT environment.
	
	\subsection{Distributed Cooperative Routing}
	Efficient inter-satellite routing is essential for maintaining reliable connectivity within the expansive and highly dynamic LEO satellite IoT. {It is worth noting that for small-scale LEO constellations, centralized routing can be a simpler and more efficient choice. A central controller can easily maintain a complete topology, making routing computations like shortest-path algorithms trivial in overhead. The absolute signaling load remains low due to the limited number of links, and the single-point-of-failure concern can often be mitigated by redundant ground stations for non-critical applications.
	However, as this work focuses on mega‑constellations with highly dynamic ISLs and limited ground visibility, the advantages of distributed cooperative routing become not only beneficial but necessary. In terms of scalability, centralized routing requires global topology collection and path computation for all node pairs, incurring $O(N^2)$ or higher complexity, whereas distributed routing relies on local neighbor exchange and achieves near‑linear scaling. For resilience, frequent link breaks and handovers render centralized schemes vulnerable to outdated tables and single points of failure, while distributed routing enables each satellite to react locally without ground coordination. Regarding signaling overhead, centralized approaches force every topology change to be reported to the controller, causing severe congestion on feeder links, while distributed schemes exchange only lightweight messages between neighbors, keeping overhead low and localized.}
	
	\begin{figure}[h]
		\centering
		\includegraphics [width=0.5\textwidth] {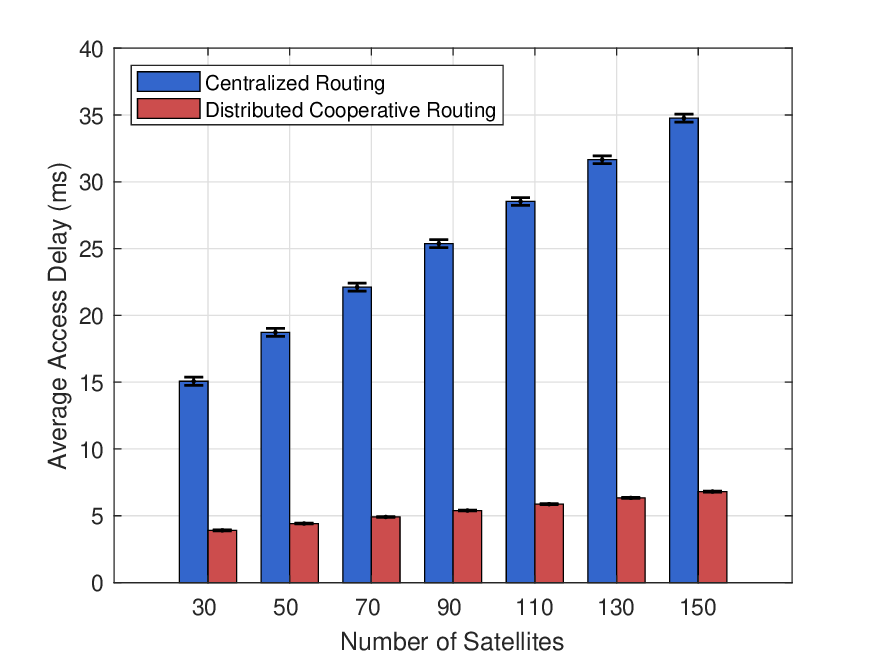}
		\caption {The average access delay performance comparisons of different routing schemes under various number of satellites.} 
		\label{Fig3-3}
	\end{figure}
	Building upon these distributed principles, recent research has introduced a distributed satellite-terrestrial cooperative routing strategy tailored for massive LEO constellations \cite{rt1}. This protocol employs a lightweight network model based on real-time orbital positions, allowing satellites to independently determine the most efficient path with minimal computational effort. By integrating ground stations as cooperative relay nodes, the system increases connectivity options and reduces overall signal delay, which preserves limited onboard memory and processing power while responding dynamically to the rapid movement of the constellation. {For inter-satellite routing, average access delay is used to evaluate latency performance under different numbers of nodes and service-area sizes and we perform 2000 trials to avoid uncertainties. The network topology consists of a variable number of LEO satellites, ranging from 30 to 150, which are uniformly and randomly distributed within a $2000 \times 2000$ km$^2$ coverage area. The maximum communication range is restricted to 1000 km.} Consequently, as demonstrated in Fig. \ref{Fig3-3}, this distributed cooperative paradigm significantly outperforms traditional centralized scheme, achieving a marked reduction in average access delay.

	\section{On-orbit Validation}	
	In this section, we present a comprehensive on-orbit verification campaign to bridge the gap between theoretical frameworks and practical deployment. Specifically, we validate the real-world feasibility of LEO satellite IoT connectivity by detailing the integrated hardware and software architectures and end-to-end transmission procedures.

	\subsection{Hardware and Software Architectures}
	The primary hardware equipment comprises Spacesail constellation (SSC), host computers (HCs), universal software radio peripheral (USRPs), radio frequency (RF) converters, and phased array antennas. The specific hardware equipment are shown as follows:
	\begin{itemize}
		\item {\textbf{SSC:} Serving as the on-orbit base stations, the SSC provides authentic space-ground links with real-world Doppler shifts and propagation delays to validate D2S communications. The selected SSC satellite used in the validation operates in a near-polar LEO orbit with an orbital altitude of approximately 800 km. Its two-line orbital element (TLE) is used to calculate the time-varying azimuth, elevation, slant range, and Doppler trend during the satellite pass, which provides the reference information for antenna pointing and link acquisition.}
	
		\item \textbf{HC}: The HC handles baseband processing and system control, utilizing an Intel Core i7-13650HX processor and a NVIDIA GeForce RTX 4060 graphics processing unit (GPU) for computational acceleration.
		\item \textbf{USRP}: The NI Ettus USRP X410 facilitates digital-to-analog/analog-to-digital conversion (DAC/ADC), supporting a frequency range of 1 MHz to 7.2 GHz, extendable to 8 GHz.
		\item \textbf{RF converter}: The RF converter performs bidirectional heterodyne conversion between the L-band and Ku-band.
		\item {\textbf{Phased array antenna:} The phased array antenna is a Ku-band two-dimensional active phased-array terminal. It supports 10.7--12.7 GHz reception and 13.7--14.5 GHz transmission, automatic polarization switching, a G/T no less than 9.5 dB/K, an EIRP no less than 41 dBW, a 5-degree narrow beamwidth, and a tracking accuracy within 0.5 dB RMS.}
	\end{itemize}
	
	The software environment integrates three main platforms. MATLAB R2025a provides comprehensive baseband processing capabilities, including channel coding, interleaving, and modulation, with GPU acceleration through the parallel computing toolbox. NI LabVIEW 2021 offers real-time control and data acquisition for the USRP, managing frequency tuning, gain control, and high-speed in-phase and quadrature (IQ) data streaming up to 491.52 MS/s. The Xphased software implements beamforming algorithms and autonomous tracking with a 50 Hz update frequency, maintaining communication link stability despite the high-speed mobility of LEO satellites. These software components communicate through standardized interfaces, ensuring synchronization for the entire system.

	\subsection{End-to-end Transmission}
\begin{figure*}[htbp]
		\centering
		\includegraphics[width=1\textwidth]{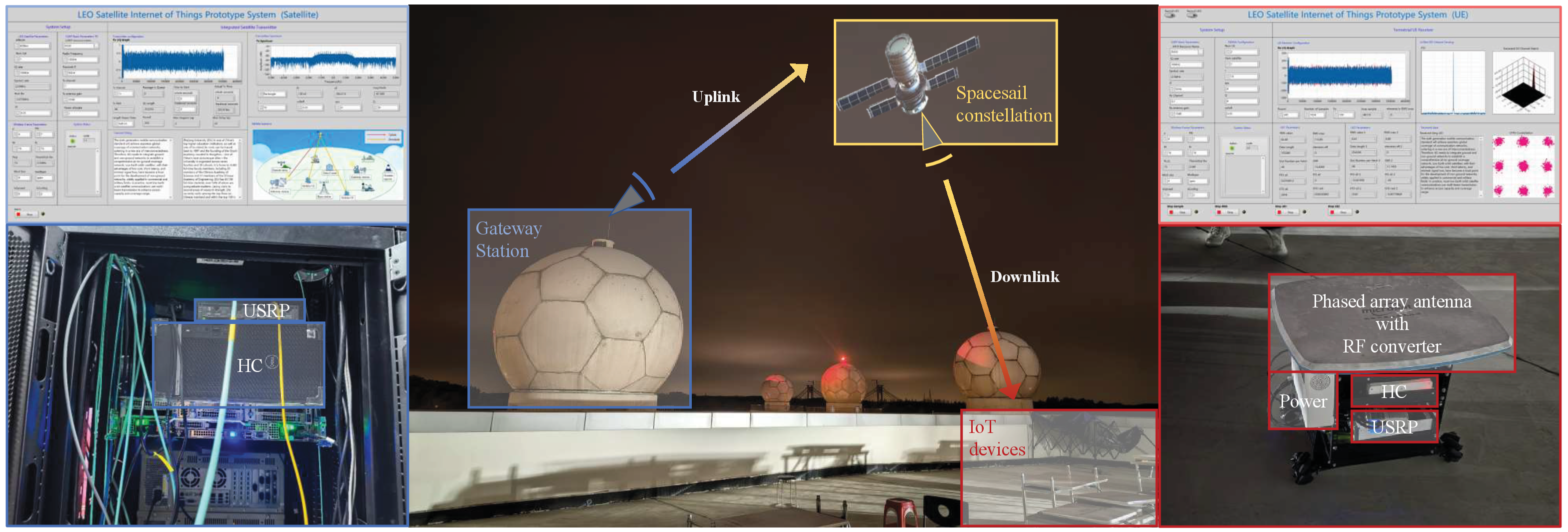}
		\caption{On-orbit validation scenario of LEO satellite IoT.}
		\label{real-scenario}
	\end{figure*}

	As shown in Fig. \ref{real-scenario}, the satellite-ground communication link is established through three primary components: the gateway station, a transparent relay LEO satellite constellation, and IoT devices.
	
	{The gateway station is equipped with an HC, a USRP, an up-converter, and an autonomous tracking antenna. Baseband communication frames adopting an OTFS waveform with LDPC channel coding are generated in MATLAB on the HC, where GPU acceleration is employed to enable parallelized IQ data synthesis and pilot sequence insertion. The pilot symbols are periodically embedded within each frame to support channel estimation, SNR estimation, and synchronization. Next, LabVIEW running on the HC delivers the generated frames and control signals to the USRP over the peripheral component interconnect express (PCIe) bus. Subsequently, the USRP performs digital-to-analog conversion and outputs the resulting analog signals in the L-band. The transmit power at the USRP output is approximately -20 dBm. Ultimately, the L-band signal undergoes amplification and up-conversion at the gateway station, resulting in a Ku-band uplink signal that is transmitted to the transparent relay LEO satellite.
	
	Upon reception, the LEO satellite processes the uplink signal through a receiving chain consisting of low-noise amplification, frequency translation, and power amplification. As a transparent relay node, the satellite does not perform baseband regeneration but instead forwards the signal after amplification. A large-scale phased array antenna onboard the satellite applies digital beamforming to steer high-gain downlink beams toward the coverage area of the IoT devices. In the on-orbit validation, the configured carrier frequencies are 14.0925 GHz with right-hand circular polarization for the uplink and 10.775 GHz with left-hand circular polarization for the downlink. A downlink beam with a maximum bandwidth of 125 MHz is used, while the USRP I/Q rate and symbol rate are set to 120 MS/s and 30 Mbaud, respectively.
	
	The IoT device is emulated by an HC, a USRP, and a phased array antenna. First, based on the published TLE and the inherent downlink parameters of the LEO constellation, the HC utilizes Xphased software to generate control commands through the open antenna management interface protocol (OPENAMIP). These commands dynamically steer the phased array antenna to receive the satellite Ku-band downlink signals, which are then accurately converted to the L-band for subsequent demodulation. Next, the L-band signal is routed to the USRP, where it is digitized into baseband I/Q data. The received SNR estimated from the m-sequence pilot is higher than 15 dB. This data is subsequently streamed via PCIe to the HC. Finally, the HC leverages GPU acceleration to process multiple transmission time slots in parallel, ensuring low-latency, real-time decoding of the received signals. The receiver successfully completes frame synchronization, OTFS demodulation, LDPC decoding, and recovery of text and image payloads, thereby validating the feasibility of the end-to-end D2S IoT link under real on-orbit dynamics.}
	
	Ultimately, the successful execution of this end-to-end transmission chain validates the real-world feasibility and effectiveness of the proposed architecture. It comprehensively demonstrates the system's capability to maintain robust LEO satellite IoT connectivity under highly dynamic environments.
	
	\section{Challenges and Future Directions}
	In this section, we identify the critical challenges and future research directions for LEO satellite IoT, including Doppler-resilient waveform design, resource-constrained onboard intelligence, and the deep integration  of communication, navigation, and remote sensing on satellite platforms. 
	
	\textbf{Effective Advanced Waveforms Design:} Designing physical-layer waveforms for LEO satellite IoT necessitates overcoming the severe Doppler shifts and doubly-selective fading induced by high mobility channels. While traditional orthogonal frequency division multiplexing struggles under these conditions, the newly proposed Delay-Doppler domain modulations, such as orthogonal time frequency space (OTFS) and orthogonal Delay-Doppler multiplexing (ODDM), offer a promising solution by transforming time-varying fading into a quasi-stationary interaction. However, the substantial computational overhead for signal detection and channel estimation in these schemes remains a significant barrier for energy-constrained IoT devices. To bridge this gap, future research must prioritize low-complexity algorithms and sparse pilot optimization to drastically reduce processing loads.
	
	\textbf{Onboard Intelligence under Resource Constraints:} The deployment of AI, particularly DL, on LEO satellites presents a critical tension between performance gains and severe platform limitations. While AI enables intelligent beamforming and autonomous routing, stringent onboard power constraints heavily restrict the complexity of deployable models. Therefore, achieving extreme model lightweighting through advanced techniques, such as neural architecture search, pruning, and quantization, remains a formidable challenge. Ultimately, advancing these lightweight AI technologies is paramount to unlocking the full potential of onboard intelligence while preserving the operational viability of resource-constrained LEO satellite IoT.

	\textbf{Integration of Communication, Navigation, and Remote Sensing:} The evolution of LEO satellite IoT necessitates integrating communication, navigation, and remote sensing to mitigate the fragmented resource utilization and high operational costs of traditional, disparate platforms. However, integrating these capabilities introduces intense competition for limited onboard power and shared spectrum, often exacerbating signal interference between high-power transmitters and sensitive receivers. Consequently, future research must prioritize integrated hardware architectures and unified waveform designs that enable simultaneous high-speed transmission, environmental observation, and precise positioning to support complex, multi-dimensional IoT services.

   \section{Conclusion} 
     {LEO satellite IoT represents a transformative shift toward universal connectivity in the 6G era. To enable this paradigm, this article proposed a holistic, two-dimensional system architecture for LEO satellite IoT. Building upon this architecture, we detailed three pivotal enabling technologies, including massive grant-free random access, deep learning-based multi-beam precoding, and distributed cooperative routing, tailored to optimize the uplink, downlink, and inter-satellite links, respectively. Furthermore, we presented an on-orbit verification platform to validate the real-world feasibility of D2S connectivity. Finally, by outlining critical open challenges and future research directions, this article serves as a robust technical reference to propel the realization of a resilient, globally ubiquitous 6G LEO satellite IoT.}


\begin{thebibliography}{1}

\bibitem{r1}
C.-X. Wang \emph{et al}., ``On the road to 6G: Visions, requirements, key technologies and testbeds,” \emph{IEEE Commun. Surveys Tuts.}, vol. 25, no. 2, pp. 905–974, 2023.

\bibitem{r2}
M. A. Ullah, K. Mikhaylov, and H. Alves, ``An overview of direct-to-satellite IoT: opportunities and open challenges,” in \emph{Proc. IEEE 9th World Forum Internet Things (WF-IoT)}, Aveiro, Portugal, Oct. 2023, pp. 1–8.

\bibitem{r3}
D.-H. Jung, G. Im, J.-G. Ryu, S. Park, H. Yu, and J. Choi, ``Satellite clustering for non-terrestrial networks: Concept, architectures, and applications,'' \emph{IEEE Veh. Technol. Mag.}, vol. 18, no. 3, pp. 29–37, Sep. 2023.

\bibitem{r4}
J. A. Fraire, O. Iova, and F. Valois, ``Space-terrestrial integrated Internet of Things: Challenges and opportunities,"\emph{IEEE Commun. Mag.}, vol. 60, no. 12, pp. 64–70, Dec. 2022.

\bibitem{r5}
T. Ma, B. Qian, X. Qin, X. Liu, H. Zhou, and L. Zhao, ``Satellite-terrestrial integrated 6G: An ultra-dense LEO networking management architecture," \emph{IEEE Wirel. Commun.}, vol. 31, no. 1, pp. 62–69, Feb. 2024.

\bibitem{r6}
T. Huang \emph{et al.}, ``Integrated computing and networking for LEO satellite mega-constellations: Architecture, challenges and open issues," \emph{IEEE Wirel. Commun.}, vol. 31, no. 5, pp. 92–100, Oct. 2024.

\bibitem{r7}
3GPP TR38.811, ``Study on new radio (NR) to support non-terrestrial networks (Release 15)," tech. rep., 3rd Generation Partnership Project (3GPP), Sep. 2020.

\bibitem{r8}
3GPP TR 38.821, ``Solutions for NR to support non-terrestrial networks (NTN) (Release 16)," tech. rep., 3rd Generation Partnership Project (3GPP), May 2021.

\bibitem{r9}
H. Shahid \emph{et al.}, ``Emerging advancements in 6G NTN radio access technologies: An overview," in \emph{Proc. Joint Eur. Conf. Netw. Commun. 6G Summit (EuCNC/6G Summit)}, Antwerp, Belgium, 2024, pp. 593–598.

\bibitem{r10}
O. Kodheli \emph{et al.}, ``Satellite communications in the new space era: A survey and future challenges," \emph{IEEE Commun. Surv. Tutor.}, vol. 23, no. 1, pp. 70–109, 1st Quart. 2021.

\bibitem{r11}
X. Luo, H.-H. Chen, and Q. Guo, ``LEO/VLEO satellite communications in 6G and beyond networks: Technologies, applications, and challenges," \emph{IEEE Netw.}, vol. 38, no. 5, pp. 273–285, Sep. 2024.

\bibitem{r12}
L. Jin, L. Wang, X. Jin, J. Zhu, K. Duan, and Z. Li, ``Research on the application of LEO satellite in IoT," in \emph{Proc. IEEE 2nd Int. Conf. Electron. Technol., Commun. Inf. (ICETCI)}, Changchun, China, 2022, pp. 739–741.

\bibitem{ma1}
M. Ying, X. Chen, and X. Shao, ``Exploiting tensor-based Bayesian learning for massive grant-Free random access in LEO satellite Internet of Things," \emph{IEEE Trans. Commun.}, vol. 71, no. 2, pp. 1141-1152, Feb. 2023.

\bibitem{dl1}
M. Ying, X. Chen, Q. Qi, and W. Gerstacker, ``Deep learning-based joint channel prediction and multibeam precoding for LEO satellite Internet of Things," \emph{IEEE Trans. Wireless Commun.}, vol. 23, no. 10, pp. 13946-13960, Oct. 2024.

\bibitem{rt1}
X. Feng, Y. Sun, and M. Peng, ``Distributed satellite-terrestrial cooperative routing strategy based on minimum hop-count analysis in mega LEO satellite constellation,'' \emph{IEEE Trans. Mobile Comput.}, vol. 23, no. 11, pp. 10678-10693, Nov. 2024.


\end{thebibliography}
\end{document}